\documentclass[aps,prd,showpacs,floatfix,preprint]{revtex4}
\usepackage{amsmath,bm}
\usepackage{graphicx}
\usepackage{epsfig}
\begin{document}

\title{Effect of Shear viscosity on the nucleation of antikaon condensed matter in neutron stars}

\author{Sarmistha Banik$^1$ and Debades Bandyopadhyay$^2$}
\affiliation{$^1$Variable Energy Cyclotron Centre,
1/AF Bidhannagar, Kolkata-700064, India}
\affiliation{$^2$Astroparticle Physics and Cosmology Division,
Saha Institute of Nuclear Physics, 1/AF Bidhannagar, 
Kolkata-700064, India}

\begin{abstract}
We investigate a first-order phase transition from hadronic matter to antikaon
condensed matter during the cooling stage of protoneutron stars. The phase 
transition proceeds through the thermal nucleation of antikaon condensed 
matter. In this connection we study the effect of shear viscosity on the 
thermal nucleation rate of droplets of antikaon condensed matter. Here we adopt
the same equation of state for the calculation of shear viscosity and thermal
nucleation time. We compute the shear viscosity 
of neutron star matter composed of neutrons, protons, electrons and muons using
the relativistic mean field model. The prefactor in the nucleation rate 
which includes the shear viscosity, is enhanced by several orders of 
magnitude compared with the $T^4$ approximation of earlier calculations. 
Consequently the thermal nucleation time in the $T^4$ approximation 
overestimates our result. Further the thermal nucleation of
an antikaon droplet might be possible in our case for surface tension smaller 
than 20 MeV fm$^{-2}$.
\pacs{97.60.Jd, 26.60.-c,52.25.Fi,64.60.Q-}
\end{abstract}

\maketitle

\section{Introduction}
Antikaon ($K^-$ meson) condensation in dense baryonic matter formed in heavy 
ion collisions as well as in neutron stars was first proposed by Kaplan and 
Nelson \cite{Kap}. There was lots of interest in the study of antikaon 
condensation in neutron stars 
\cite{Bro,Tho,Ell,Lee,Pra97,Gle99,Kno,Sch,Pal,Bani1,Bani2,Bani3,Bani4,Pons,Bani5}
after their work. A first-order phase transition from hadronic matter to 
antikaon condensed matter was investigated in several cases 
using relativistic field theoretical models \cite{Gle99,Bani1,Pons,Bani5}. The
phase transition was either studied using Maxwell construction or governed by 
Gibbs' rules for phase equilibrium along with global baryon number and charge 
conservation \cite{Gle92}. In those cases the focus was on the equation of 
state (EoS) and neutron star structure as well as critical temperature of 
antikaon condensation. 

A first-order phase transition may proceed through the nucleation of droplets
of the new phase. The formation of droplets of exotic matter such as antikaon 
condensed matter and quark matter, could be possible in neutron stars when the 
protoneutron star cools down to a temperature of $\sim$ 10 MeV and is 
deleptonised
\cite{Nor,San,Sat,Hei,Bom,Min}. This nucleation process could be due to quantum
and
thermal nucleation mechanisms \cite{Bom}. The thermal nucleation of antikaon 
condensed matter was already studied using the homogeneous nucleation theory of
Langer \cite{Nor,San,Lan}. Recently it has been shown in the context of 
nucleation of quark matter in hot and neutrino-free neutron stars that the 
thermal nucleation is more efficient than the quantum nucleation process at 
higher temperatures \cite{Bom}.   

The homogeneous nucleation theory of Langer \cite{Lan,Tur} is applicable
close to a first-order phase transition. The hadronic matter becomes metastable
near the phase transition point when there is a sudden change in state 
variables. In this case thermal and quantum fluctuations play important roles
in the metastable phase. Small ranged and localised fluctuations in state 
variables of the metastable hadronic matter might lead to the nucleation of 
droplets of stable antikaon condensed matter. Those droplets of the stable 
phase having radii larger than a critical radius will survive and grow. A  
droplet may grow beyond the critical size if the latent heat is transported 
from the surface of the droplet into the metastable phase. It was argued that
the heat transportation could be achieved through the thermal dissipation and
viscous damping \cite{Tur,Las,Raj}. The influence of thermal conductivity and
shear viscosity on the thermal nucleation time was studied in a first-order
phase transition from hadronic to quark matter \cite{Bom,Las}. However there is
no such calculation for the thermal nucleation of antikaon condensed matter.

We are motivated to study the effect of shear viscosity on the thermal 
nucleation rate of droplets of antikaon condensed matter in this work. Shear
viscosity of pure neutron and neutron star matter has been calculated by 
several 
groups \cite{Flo1,Flo2,Cut,Ben,Yak,Glam}. Recently we have investigated 
the shear viscosity in antikaon condensed matter 
\cite{Bani6}. Though we considered a first-order antikaon condensation, we did
not take into account the nucleation process. We performed the calculation of 
shear viscosity in neutron star matter using the EoS derived from relativistic 
field theoretical models. Now the question is how this shear 
viscosity of nucleonic phase may impact the nucleation of antikaon condensed 
phase. The earlier calculation of the nucleation of quark matter adopted a 
parametrised form of the shear viscosity \cite{Bom}. In this calculation we 
adopt the same EoS for the computation of thermal nucleation time
of the antikaon condensed phase and shear viscosity of nuclear matter.    

We organise the paper in the following way. We describe models for 
homogeneous nucleation, shear viscosity and EoS in Sec. II.
Results of this calculation are discussed in Sec. III. Sec. IV gives the
summary and conclusions.

\section{Formalism}
We consider a first-order phase transition from the charge neutral and 
beta-equilibrated nuclear matter to $K^-$ condensed matter in a hot neutron 
star after the emission of trapped neutrinos. Droplets of antikaon condensed
phase are formed in the metastable nuclear matter due to thermal fluctuations.
Droplets of antikaon condensed matter above a critical size are of interest
because those critical droplets will drive the phase transition. We adopt the
homogeneous nucleation formalism of Langer to calculate the thermal nucleation
rate \cite{Lan}. In this formalism, the thermal nucleation per unit time per 
unit volume is given by \cite{Lan,Tur}
\begin{equation}
\Gamma = \Gamma_0 exp\left(-\frac {\triangle F (R_c)}{T}\right)~,
\end{equation}  
where $\triangle F$ is the change in the free energy to produce a critical 
droplet in the metastable nuclear matter. The change in free energy of the
system due to the formation of a droplet is given by \cite{Hei,Min}
\begin{equation}
\triangle F (R) = -\frac{4\pi}{3} (P^K - P^N) R^3 + 4\pi \sigma R^2~,
\end{equation}
where $R$ is the radius of the droplet, $\sigma$ is surface tension of the
interface separating two phases and $P^N$ and $P^K$ are the pressure in 
nuclear and antikaon condensed phases respectively. We obtain the critical  
radius of the droplet from the maximum of $\triangle F(R)$ i.e. 
$\delta_R \triangle F = 0$ and it is 
\begin{equation}
R_C = \frac{2 \sigma}{(P^K - P^N)}~.
\end{equation}
The EoS enters into this calculation through the difference in 
pressures of two phases. 

The prefactor in Eq. (1) is factorised in the following way \cite{Tur,Las,Raj}
\begin{equation}
\Gamma_0 = \frac{\kappa}{2\pi} \Omega_0~. 
\end{equation}
The statistical prefactor ($\Omega_0$) gives the available phase space around 
the saddle point at $R_C$ during the passage of the droplet through it.
The expression for the statistical prefactor is 
\begin{equation}
\Omega_0 = \frac{2}{3\sqrt{3}} \left(\frac {\sigma}{T}\right)^{3/2} 
\left(\frac {R_C}{\xi}\right)^4~,
\end{equation}
Here $\xi$ is the correlation length for kaons. This correlation length is 
considered to be the width of
the interface between nuclear and antikaon condensed matter. Next we 
focus on the dynamical prefactor $\kappa$. This prefactor is responsible for
the initial exponential growth rate of a critical droplet. It is given by 
\cite{Las,Raj}
\begin{equation}
\kappa = \frac {2\sigma}{R_C^3 (\triangle w)^2} \left[ \lambda T + 2 
(\frac{4}{3} \eta + \zeta)\right]~.
\end{equation}
Here $\triangle w = w_{K} - w_{N}$ is the enthalpy difference of two phases
whereas $\lambda$ is thermal conductivity and $\eta$ and $\zeta$ are 
the shear and bulk viscosities of nuclear matter. It was shown that 
the latent heat would be taken away from the surface of the droplet due 
thermal dissipation and viscous damping \cite{Las,Raj}. 

Finally the thermal nucleation time ($\tau_{th}$) in the interior of neutron 
stars is calculated as
\begin{equation}
\tau_{th} = \left(V \Gamma \right)^{-1}~,
\end{equation}
where the volume $V = 4{\pi}/3 R_{nuc}^3$. We assume that pressure and
temperature are constant within this volume in the core. It is evident from 
Eq. (7) that the thermal nucleation time is inversely proportional to transport
coefficients. In connection with the quark matter nucleation, it was noted that
the contribution of thermal conductivity is not appreciable. Further, we know 
that shear viscosity increases \cite{Bani6} and bulk viscosity decreases 
\cite{Chat} as temperature decreases.  After the emission of neutrinos a hot
neutron star cools down, and the thermal nucleation drives the phase transition
when the temperature is $\sim 10$ MeV. In this temperature regime, shear 
viscosity might be an important factor.

Now we focus on the calculation of shear viscosity in neutron star matter.  
It was noted that the main contributions to the total shear viscosity in 
neutron star matter came from electrons, the lightest charged particles, and 
neutrons, the most abundant particles \cite{Flo1,Flo2,Cut,Yak,Glam}. Recently
the proton shear viscosity, though small, was considered in our calculation
\cite{Bani6}. We calculate shear viscosities for different particle species 
using coupled Boltzmann transport equations \cite{Flo2,Yak,Bani6}. 
The effects of the exchange of transverse plasmons in the collisions of charged
particles are included in this case \cite{Yak,Yak2}. The knowledge of 
nucleon-nucleon scattering cross sections derived in the Dirac-Brueckner 
approach \cite{Mach1,Mach2} is used for the calculation of neutron and proton 
shear viscosities \cite{Yak,Bani6,Bai}. Contributions of neutrons, protons, 
electrons and muons to the total shear viscosity 
\begin{equation}
\eta_{total} = \eta_n + \eta_p + \eta_e + \eta_{\mu}~,
\end{equation}
are given by \cite{Yak,Bani6}
\begin{equation}
\eta_{i(=n,p,e,\mu)} = \frac {n_i p_{F_i}^2 \tau_i}{5m_i^*}~.
\end{equation}
Here $\tau_i$ is the relaxation time of i-th species as calculated in Ref. 
\cite{Yak,Bani6}. Effective mass and Fermi momentum of i-th particle
species are denoted by $m_i^*$ and $p_{F_i}$, respectively. For electrons and 
muons, 
effective masses are taken as their chemical potentials due to relativistic 
effects. We adopt a relativistic field theoretical model which is described in
the following paragraphs, for the calculation effective masses and Fermi 
momenta of neutrons and protons. 

Besides the computation of shear viscosities, the knowledge of the EoS 
is essential for the thermal nucleation rate. We note that the EoS enters into 
the calculation of thermal nucleation rate as given by Eqs. (1)-(3). 
We adopt relativistic field theoretical models to describe the 
$\beta$ equilibrated matter in nuclear and antikaon condensed phases. Those two 
phases are composed of neutrons,
protons, electrons, muons and of $K^-$ mesons only in the antikaon condensed
phase. Both phases are governed by baryon number conservation and charge 
neutrality conditions \cite{Gle92}. The baryon-baryon interaction mediated by
the exchange of $\sigma$, $\omega$ and $\rho$ mesons is described by the
Lagrangian density \cite{Sch,walecka,serot,glendenning}    
\begin{eqnarray}
{\cal L}_N &=& \sum_{B=n,p} \bar\psi_{B}\left(i\gamma_\mu 
{\partial}^{\mu} - m_B
+ g_{\sigma B} \sigma - g_{\omega B} \gamma_\mu \omega^\mu
- g_{\rho B}
\gamma_\mu{\mbox{\boldmath t}}_B \cdot
{\mbox{\boldmath $\rho$}}^\mu \right)\psi_B\nonumber\\
&& + \frac{1}{2}\left( \partial_\mu \sigma\partial^\mu \sigma
- m_\sigma^2 \sigma^2\right) - U(\sigma) \nonumber\\
&& -\frac{1}{4} \omega_{\mu\nu}\omega^{\mu\nu}
+\frac{1}{2}m_\omega^2 \omega_\mu \omega^\mu
- \frac{1}{4}{\mbox {\boldmath $\rho$}}_{\mu\nu} \cdot
{\mbox {\boldmath $\rho$}}^{\mu\nu}
+ \frac{1}{2}m_\rho^2 {\mbox {\boldmath $\rho$}}_\mu \cdot
{\mbox {\boldmath $\rho$}}^\mu ~.
\end{eqnarray}
The scalar self-interaction term \cite{Sch,glendenning,boguta} is 
\begin{equation}
U(\sigma)~=~\frac13~g_1~m_N~(g_{\sigma N}\sigma)^3~+~ \frac14~g_2~
(g_{\sigma N}\sigma)^4~,
\end{equation}
The effective nucleon mass is given by $m_B^* = m_B - g_{\sigma B} \sigma$, 
where $m_B$ is the vacuum baryon mass.
The Lagrangian density for (anti)kaons in the minimal coupling is given by
\cite{Pal,Bani2,Gle99}
\begin{equation}
{\cal L}_K = D^*_\mu{\bar K} D^\mu K - m_K^{* 2} {\bar K} K ~,
\end{equation}
where the covariant derivative is
$D_\mu = \partial_\mu + ig_{\omega K}{\omega_\mu}
+ i g_{\rho K}
{\mbox{\boldmath t}}_K \cdot {\mbox{\boldmath $\rho$}}_\mu$ and
the effective mass of (anti)kaons is
$m_K^* = m_K - g_{\sigma K} \sigma$.
The in-medium energy of $K^{-}$ mesons is given by
\begin{equation}
\omega_{K^{-}} =  \sqrt {(p^2 + m_K^{*2})} - \left( g_{\omega K} \omega_0
+ \frac {1}{2} g_{\rho K} \rho_{03} \right)~.
\end{equation}
We consider the s-wave (${\bf p}=0$) condensation. The condensation sets in 
when the chemical potential of $K^-$ mesons  
($\mu_{K^-} = \omega_{K^-}$) is equal to the electron chemical potential i.e. 
$\mu_e = \mu_{K^-}$. The critical droplet of antikaon condensed matter is in 
total phase equilibrium with the metastable nuclear matter. The mixed phase is 
governed by Gibbs' phase rule along with global baryon number conservation
and charge neutrality \cite{Gle92}. Two phases are in chemical equilibrium and 
the mechanical equilibrium is constrained by Eq. (3). 

We do not consider the variation of the meson fields in the droplet with 
position. This is known as the bulk approximation \cite{San}. We solve 
equations of motion self-consistently in the
mean field approximation \cite{walecka} and find effective masses and Fermi 
momenta of baryons. Further we obtain the pressure in nuclear matter ($P^N$) 
and in antikaon condensed matter ($P^K$) as given by Ref.\cite{Bani2}. 

\section{Results and Discussions}
We adopt the Glendenning and Moszkowski parameter set known as GM1 
\cite{Gle91} for nucleon-meson coupling 
constants which are obtained by reproducing the saturation properties of
nuclear matter such as binding energy $E/B=-16.3$ MeV, baryon density 
$n_0=0.153$ fm$^{-3}$, asymmetry energy coefficient $a_{\rm asy}=32.5$ MeV, 
incompressibility $K=300$ MeV and effective nucleon mass $m^*_N/m_N = 0.70$. 
Next we determine kaon-meson vector coupling constants using
the quark model and isospin counting rule \cite{Gle99,Bani6}. And the scalar 
kaon-meson coupling constant is obtained from the real part of $K^-$ optical 
potential depth at normal nuclear matter density.

Heavy ion collision experiments and $K^-$ atomic data indicated an attractive
potential for antikaons and a repulsive potential for kaons 
\cite{Fri94,Fri99,Koc,Waa,Li,Pal2}. However the strength of antikaon optical
potential depth is a debatable issue. It was found from the analysis of $K^-$ 
atomic data that the real part of the antikaon optical potential could be as 
large as $U_{\bar K} = -180 \pm 20$ MeV at normal nuclear matter density
\cite{Fri94,Fri99}. But theoretical models including chirally motivated coupled
channel models as well as double pole structure of $\Lambda (1405)$ could not
find such a strongly attractive antikaon optical potential depth 
\cite{Ram,Koch,Mag,Hyo}. In this calculation we take an antikaon optical 
potential depth of $U_{\bar K} = -160$ MeV at normal nuclear matter density and 
obtain the kaon-scalar meson coupling constant $g_{\sigma K} = 2.9937$. 

Here we study the thermal nucleation of antikaon condensed phase after 
the emission of trapped neutrinos followed by the evolution of a hot neutron 
star to the cold star. The stellar matter is heated during the diffusion of 
trapped 
neutrinos leading to the deleptonisation and maximum entropy per baryon $s=2$. 
It would take about a few $10^2$ s for the maximally heated star to cool down 
to a temperature 
$\sim 1$ MeV. The threshold for the appearance of antikaon condensate would
be reduced after the deleptonisation. We demonstrate this evolution of a  
neutrino-free neutron star with a few snap-shots corresponding
to entropy per baryon $s = 0, 1,$ and $2$. We adopt the finite temperature
calculation of Ref.\cite{Bani5} in this case. We show the temperature as a 
function of baryon density in nuclear matter for fixed $s=1$ and $2$ in Fig. 1.
The temperature increases with baryon density. This shows that the
temperature varies from a higher value in the core to a smaller 
value at the surface of a neutron star. The equations of state of neutrino free
nuclear
matter at fixed $s=0, 1,$ and $2$ are exhibited in Fig. 2. It is evident from 
Fig. 2 that the temperature of a few tens of MeV does not modify the 
EoS appreciably upto energy densities $\sim$ 1000 MeV fm$^{-3}$ relevant at the
center of a neutron star, with respect to the zero temperature EoS corresponding
to the $s=0$, case because the temperature is much less than the baryon 
chemical potential. Therefore,
we consider the zero temperature EoS for the rest of our calculation.
 
The total shear viscosity in nuclear matter is shown as a function of normalised
baryon density for different temperatures in Fig. 3. The shear viscosity 
decreases as temperature increases. Further, the shear viscosity increases with
increasing baryon density. It was earlier noted that the temperature
dependence of shear viscosities corresponding to different particle species was
a complicated one and manifested through relaxation times \cite{Bani6}. This 
temperature dependence deviates from the characteristic temperature 
dependence, $1/T^2$, of a Fermi liquid \cite{Yak}. Figure 3 shows
the total shear viscosity for three temperatures $T =$ 1, 10, and 30 MeV. We
are interested in the temperature range from a few MeV to a few tens of MeV in
this calculation.  

Next our focus is on the prefactor ($\Gamma_0$) of Eq. (4) which not only 
involves transport properties of nuclear matter such as thermal conductivity
and shear and bulk viscosities but also depends on the correlation length of
kaons and surface tension.
The radius of the droplet is to be greater than the correlation length for
kaons $\xi$. Otherwise many approximations in this calculation would collapse 
\cite{San,Las}. The correlation length is also taken as the thickness of the
interface between nuclear and kaon phases \cite{Bom,Las} estimated to be 
$\sim$ 5 fm \cite{San}. 
Here we consider antikaon droplets with radii greater than 5 fm. The 
other important parameter in the prefactor is the surface tension. The surface
tension between nuclear and kaon phases was already calculated by Christiansen 
and collaborators \cite{Chri}. They found that the surface tension varied from
a value of 30 MeV fm$^{-2}$ at the start of the mixed phase to a value of 10 
MeV 
fm$^{-2}$ at the end of the mixed phase. It is evident from the fit of 
Ref.\cite{Chri} that the surface tension is sensitive to the EoS. It is to be
noted that we have a different parameter set and stiffer EoS than those of 
Ref.\cite{Chri}. We perform this calculation for a set of values of surface 
tension $\sigma =$ 10, 15 and 20 MeV fm$^{-2}$. Further, we approximate the 
difference in enthalpy densities ($\triangle w$) in Eq. (6) as the difference 
of energy densities because the pressure difference is negligible compared with
the difference in energy densities \cite{Las}.       

The prefactor ($\Gamma_0$) is shown as a function of temperature in Fig. 4. 
The prefactor is calculated at a baryon density $n_b = 2.3n_0$ and surface 
tension $\sigma = 10$ MeV fm$^{-2}$. The density chosen here is just above the 
critical density for the transition. We calculate the prefactor for two values 
of correlation length $\xi=$2 and 5 fm. Solid lines correspond to the 
calculation of prefactor including the shear viscosity term and neglecting the 
thermal conductivity and bulk viscosity
terms in Eq. (6). On the other hand, dashed lines imply the prefactor including
only the thermal conductivity term. The thermal conductivity is calculated  
according to Ref.\cite{Bai}. The prefactor involving the shear viscosity term 
is a few orders of magnitude higher than that of the thermal conductivity for 
both values of kaon correlation length. It is worth mentioning here
that the contributions of thermal conductivity and bulk viscosity in the
prefactor were neglected in an earlier calculation of quark matter nucleation 
\cite{Bom}. We observe that the prefactor for $\xi=$5 fm is substantially 
reduced from that of $\xi=$2 fm in both cases. In the following paragraphs we 
discuss the results of our calculation performed with the prefactor involving 
only the shear viscosity term.

The prefactor of Eq. (4) was approximated by $T^4$ 
according to the dimensional analysis in some calculations \cite{Las,Min}. This
approximation might
differ considerably from the actual calculation of the prefactor.
Fig. 5 shows the prefactor of Eq. (4) including only 
the contribution of shear viscosity along with the prefactor approximated by
$T^4$ as a function of temperature. It is evident from Figure 5 that the 
approximated prefactor is several orders of magnitude smaller than the prefactor
calculated according to Eq. (4) at lower temperatures. However, this difference
between two calculations is reduced appreciably at higher temperatures. It 
remains to be seen how two calculations of the prefactor influence the thermal
nucleation time.

Now we discuss the nucleation time of a critical droplet of antikaon condensed 
matter calculated with the prefactor of Eq. (4) involving only the shear 
viscosity contribution. The critical radius of the droplet is obtained from the
relation in 
Eq. (3). The droplet which overcomes the maximum height of the energy barrier of
Eq. (1) occurring at the critical radius, grows exponentially. Such a droplet 
triggers the onset of antikaon condensation in the interior of a neutron star.
For surface tension $\sigma =$ 10 MeV fm$^{-2}$, we obtain an antikaon droplet
with critical radius 6.32 fm at baryon density 2.3$n_0$. This density 
is just above the critical density for transition from nuclear matter to the
droplet of antikaon condensed matter where the volume fraction of the antikaon 
condensed phase is approaching zero. We perform the calculation of thermal
nucleation of the critical droplet within a volume with $R_{nuc}=100$ m in the 
core of a neutron star as mentioned in Eq. (7). It is assumed that density, 
pressure and temperature are 
constant within this volume. The thermal nucleation time is plotted with 
temperature in Fig. 6. Results shown here are calculated with the kaon 
correlation length $\xi =$ 5 fm. For the above-mentioned case, the nucleation 
time of the critical droplet drops as temperature increases. For the case of 
$\sigma =$ 10 MeV fm$^{-2}$, the nucleation time is 1 s at a 
temperature of $T \sim$10 MeV. We also estimate thermal nucleation times of
critical antikaon droplets for two other values of surface tension $\sigma=$ 15
and 20 MeV fm$^{-2}$. It is noted that the size of the critical droplet 
increases with increasing surface tension. Radii of the critical droplets are 
9.48 and 12.6 fm corresponding to $\sigma=$ 15 and 20 MeV fm$^{-2}$, 
respectively, at a density 2.3$n_0$. We find a similar trend of the nucleation 
time with temperature
in these cases. However, the temperature corresponding to a particular 
nucleation time for example, 1 s, increases as the surface tension increases. 
It is worth mentioning here that the condensate might melt if the temperature 
is higher than the critical temperature \cite{Bani5}. We compare thermal 
nucleation times corresponding to different values of the surface tension with 
the cooling time of a protoneutron star $t_{cool} \sim$ a few $10^2$ s. 
Thermal 
nucleation of the antikaon condensed phase may be possible when the thermal 
nucleation time is less than the cooling time. It is evident from Fig. 6 that
the thermal nucleation time is strongly dependent on the surface tension. 
Thermal nucleation of an antikaon droplet is a possibility for surface tension 
$\sigma <$ 20 MeV fm$^{-2}$ because in this case the thermal nucleation time is
less than the cooling time at a temperature which might be below the critical 
temperature of antikaon condensation \cite{Bani5}. We repeat
the calculation of the thermal nucleation time with the correlation length
$\xi=$2 fm and find that the thermal nucleation time is slightly higher than 
that of $\xi=$5 fm over the whole temperature regime considered here.   

We compare the results of thermal nucleation time taking into account the 
effect of shear viscosity in the prefactor with that of the prefactor 
approximated by $T^4$. These results are shown in Fig. 7 for surface tension 
$\sigma =$ 10 MeV fm$^{-2}$ and at a density $n_b =$ 2.30$n_0$ . 
The results of the $T^4$ approximation are a few orders of magnitude higher 
than those of our calculation. For example, the thermal nucleation time is 1 s 
at
$T \sim$ 10 MeV in our case, whereas it is $\sim 10^{3}$ s at the same 
temperature for the $T^4$ approximated case. These results demonstrate the 
importance 
of including the shear viscosity in the prefactor of Eq. (4) in  the 
calculation of thermal nucleation time.

\section{Summary and Conclusions}
We have investigated the first-order phase transition from nuclear matter to
antikaon condensed matter through the thermal nucleation of a critical droplet
of anikaon condensed matter. Our main focus in this calculation is the role
of shear viscosity in the prefactor and its consequences on the thermal 
nucleation rate. We adopt the same EoS derived in the relativistic mean field
model for the calculation of shear viscosity and thermal nucleation rate. In 
this connection, we have constructed critical droplets of
antikaon condensed matter above the critical density for different values of
surface tension. Droplet radii increase with increasing surface tension. We
obtain thermal nucleation time as a function of temperature for a set of values
of surface tension and find that the thermal nucleation time is strongly 
dependent on the surface tension. Our results show that thermal nucleation 
of a critical antikaon droplet could be possible during the cooling stage of
a hot neutron star after neutrinos are emitted for a lower value of surface 
tension $\sigma <$ 20 MeV fm$^{-2}$. Further, a comparison of our results with
that of the $T^4$ approximation shows that the $T^4$ approximation 
overestimates 
our results of thermal nucleation time. This comparison highlights the 
importance of shear 
viscosity in our calculation.       

We have considered the thermal nucleation in a nonrotating neutron star and
neglected the bulk viscosity in the prefactor. 
However neutron stars rotate very fast at birth. They emit
gravitational waves and become r-mode unstable \cite{Chat}. The bulk
viscosity plays an important role in damping the r-mode instability. The bulk
viscosity might dominate over the shear viscosity above 10 MeV \cite{Chat}. It 
would be interesting to investigate the effect of bulk viscosity on the thermal 
nucleation time in a rotating neutron star along with that of shear 
viscosity.  
   
\section{Acknowledgement}
S.B. thanks the Alexander von Humboldt Foundation for the support during her 
visit at the Frankfurt Institute for Advanced Studies (FIAS).

\newpage 
\vspace{-2cm}

{\centerline{
\epsfxsize=12cm
\epsfysize=14cm
\epsffile{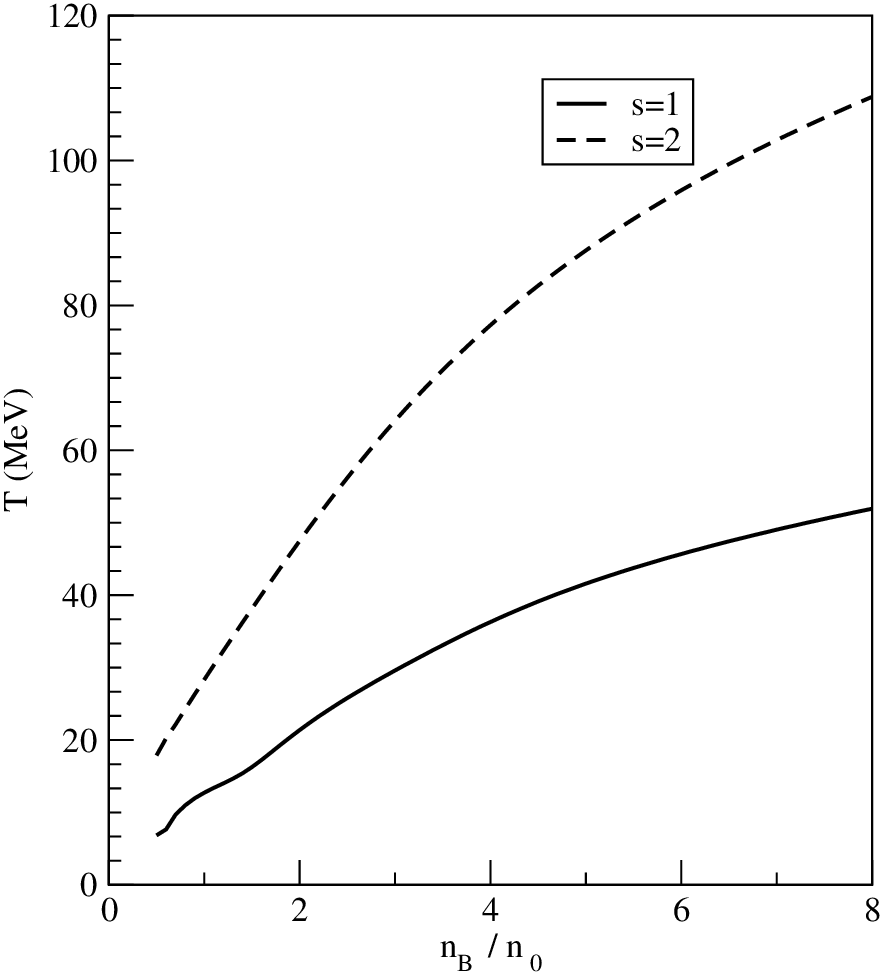}
}}

\vspace{4.0cm}

\noindent{\small{
FIG. 1. Temperature is plotted as a function of baryon density for 
neutrino-free nuclear matter and fixed entropy per baryon $s=1$ and 
$2$.}}

\newpage 
\vspace{-2cm}

{\centerline{
\epsfxsize=12cm
\epsfysize=14cm
\epsffile{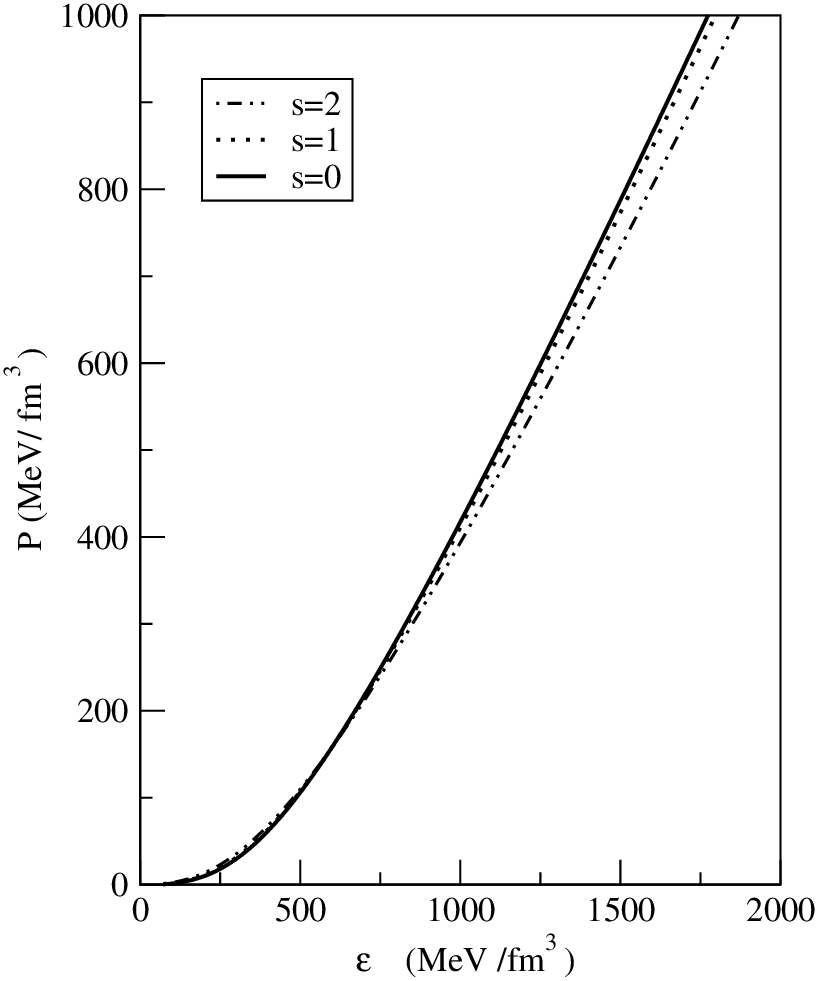}
}}

\vspace{4.0cm}

\noindent{\small{
FIG. 2. Pressure is plotted with energy density for neutrino-free nuclear
matter and fixed entropy per baryon  $s=0, 1,$ and $2$.}}

\newpage 
\vspace{-2cm}

{\centerline{
\epsfxsize=12cm
\epsfysize=14cm
\epsffile{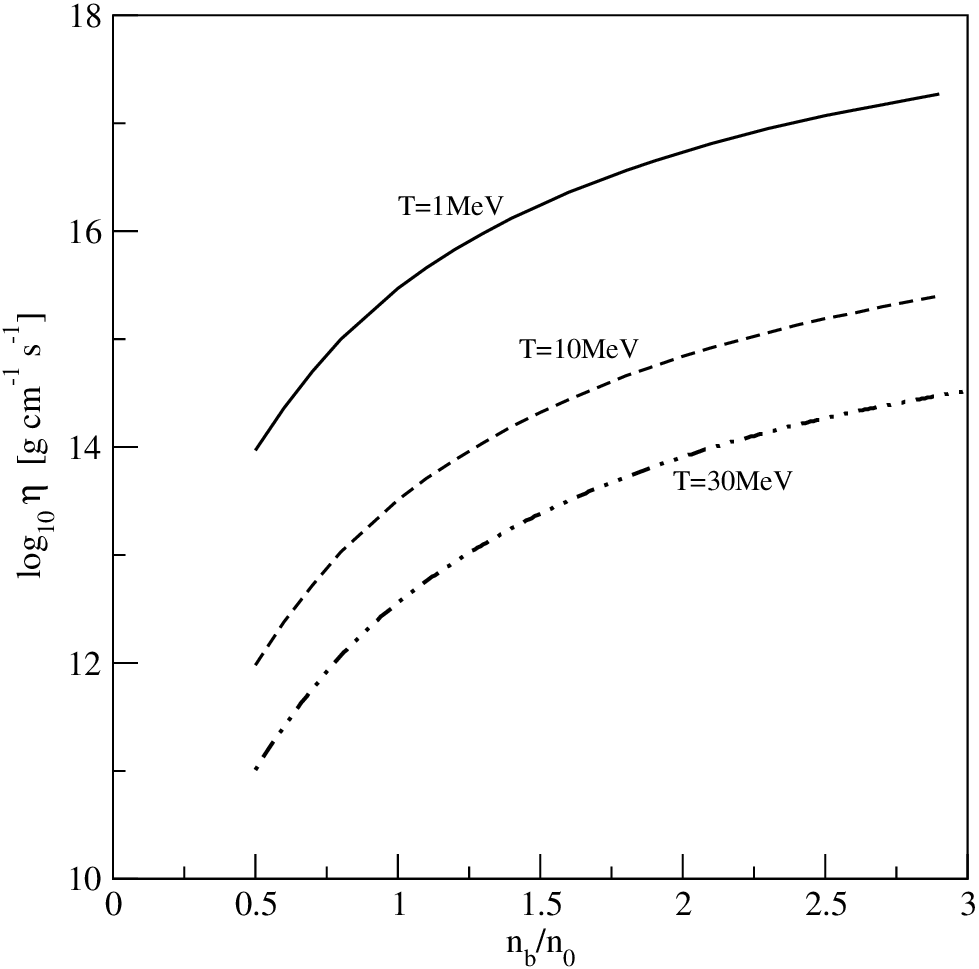}
}}

\vspace{4.0cm}

\noindent{\small{
FIG. 3. Total shear viscosity in the nuclear matter phase is plotted with 
normalised baryon density for different temperatures.}}

\newpage 
\vspace{-2cm}

{\centerline{
\epsfxsize=12cm
\epsfysize=14cm
\epsffile{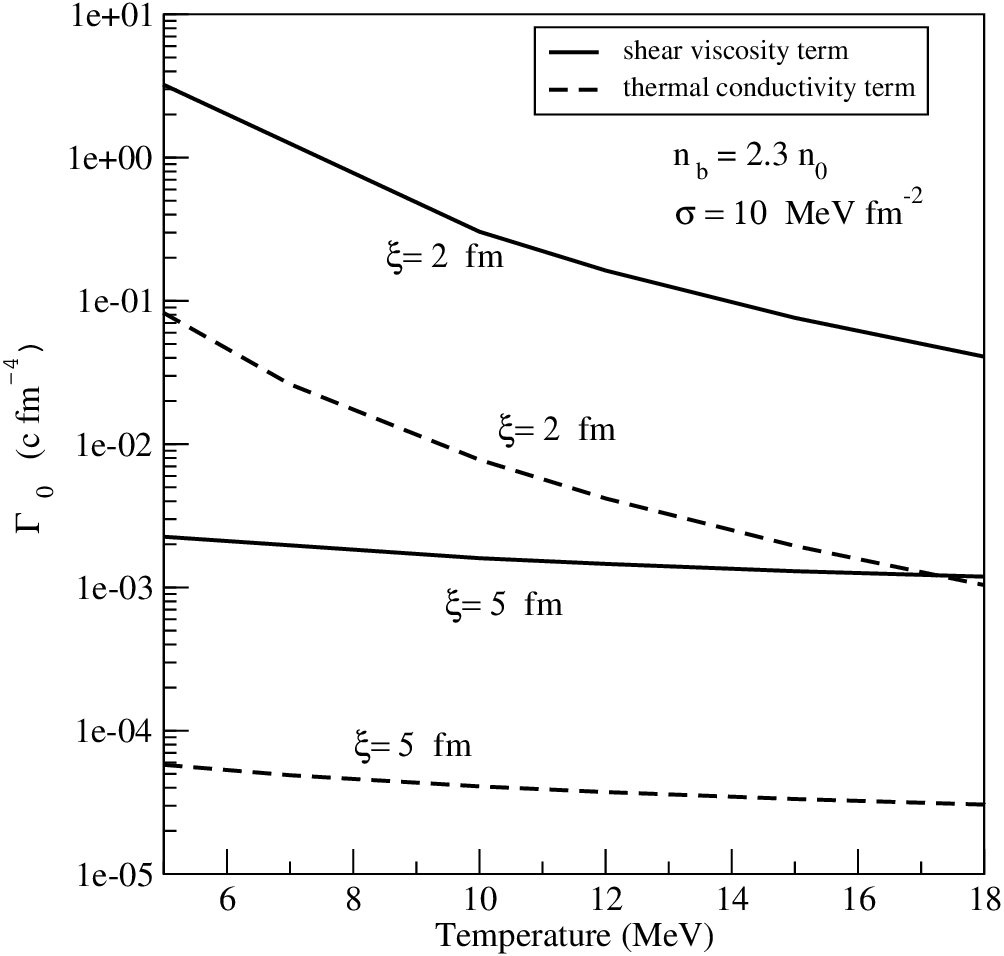}
}}

\vspace{4.0cm}

\noindent{\small{
FIG. 4. Prefactor is shown as a function of temperature for different 
correlation lengths and at a fixed density and surface tension.}}

\newpage 
\vspace{-2cm}

{\centerline{
\epsfxsize=12cm
\epsfysize=14cm
\epsffile{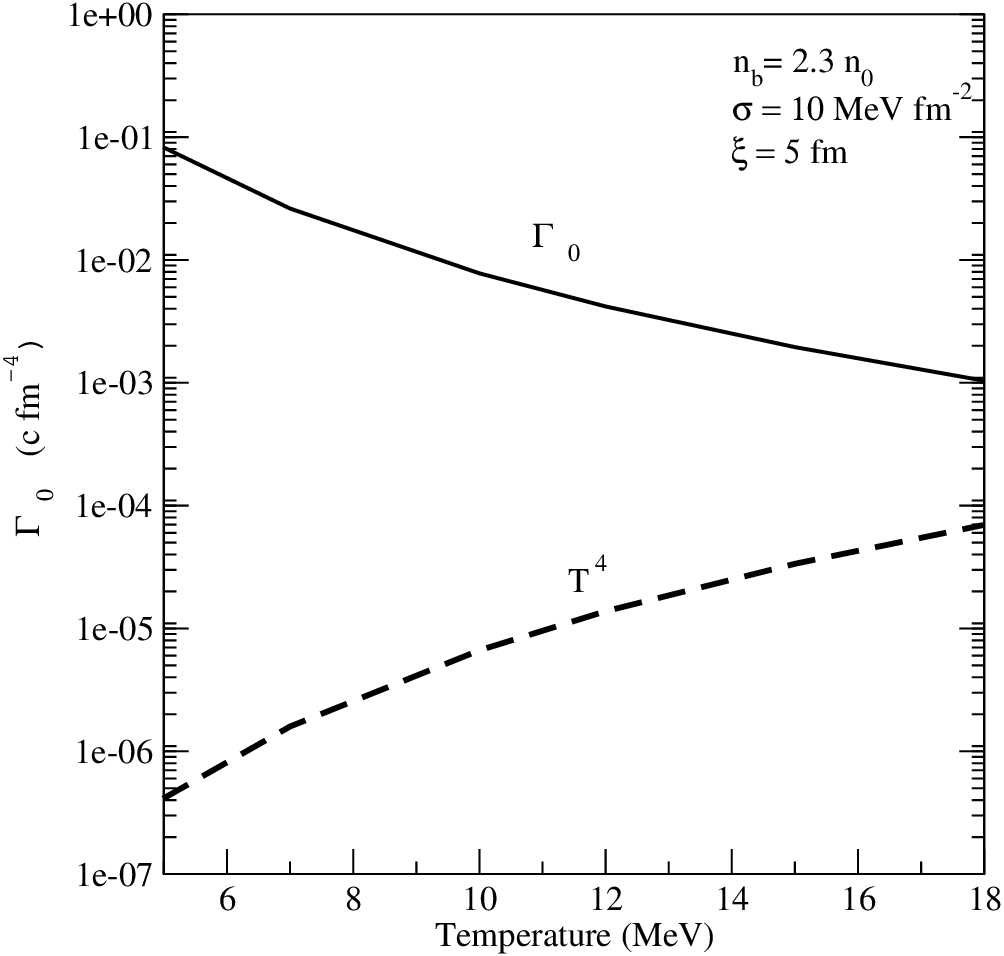}
}}

\vspace{4.0cm}

\noindent{\small{
FIG. 5. Same as Fig. 4, but the prefactor including the contribution of shear
viscosity is compared with that of $T^4$ approximation.}}

\newpage 
\vspace{-2cm}

{\centerline{
\epsfxsize=12cm
\epsfysize=14cm
\epsffile{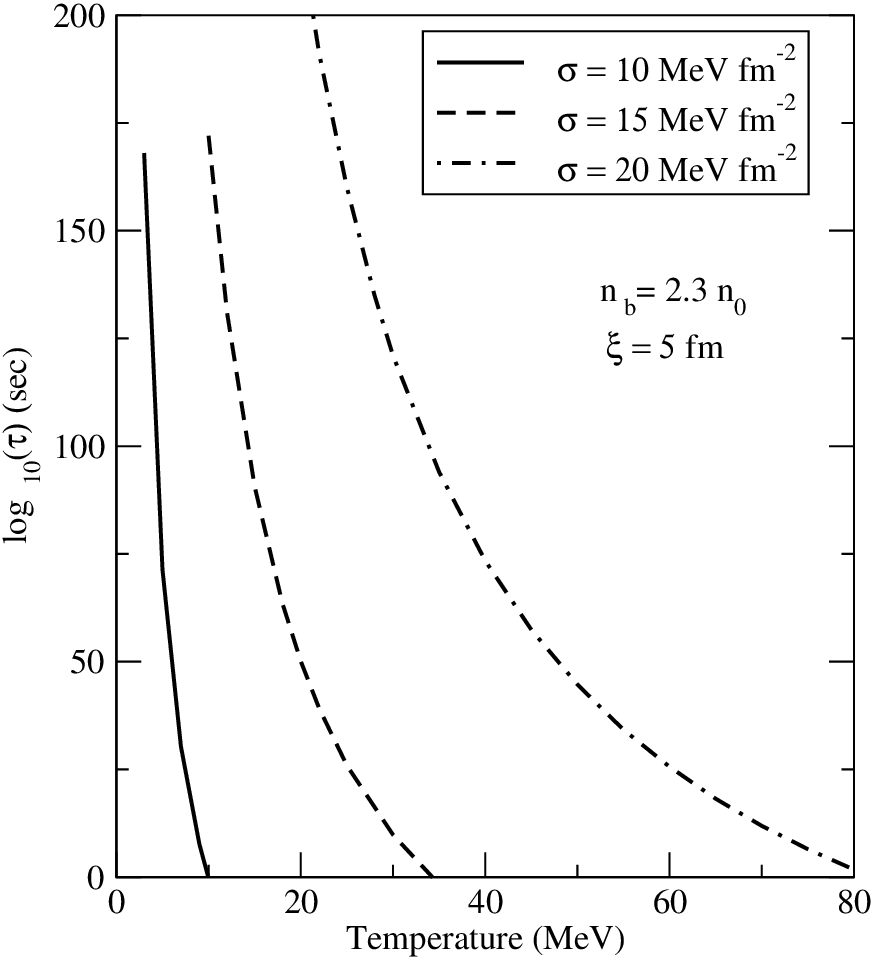}
}}

\vspace{4.0cm}

\noindent{\small{
FIG. 6. Thermal nucleation time is displayed with temperature for different
values of surface tension.}}

\newpage 
\vspace{-2cm}

{\centerline{
\epsfxsize=12cm
\epsfysize=14cm
\epsffile{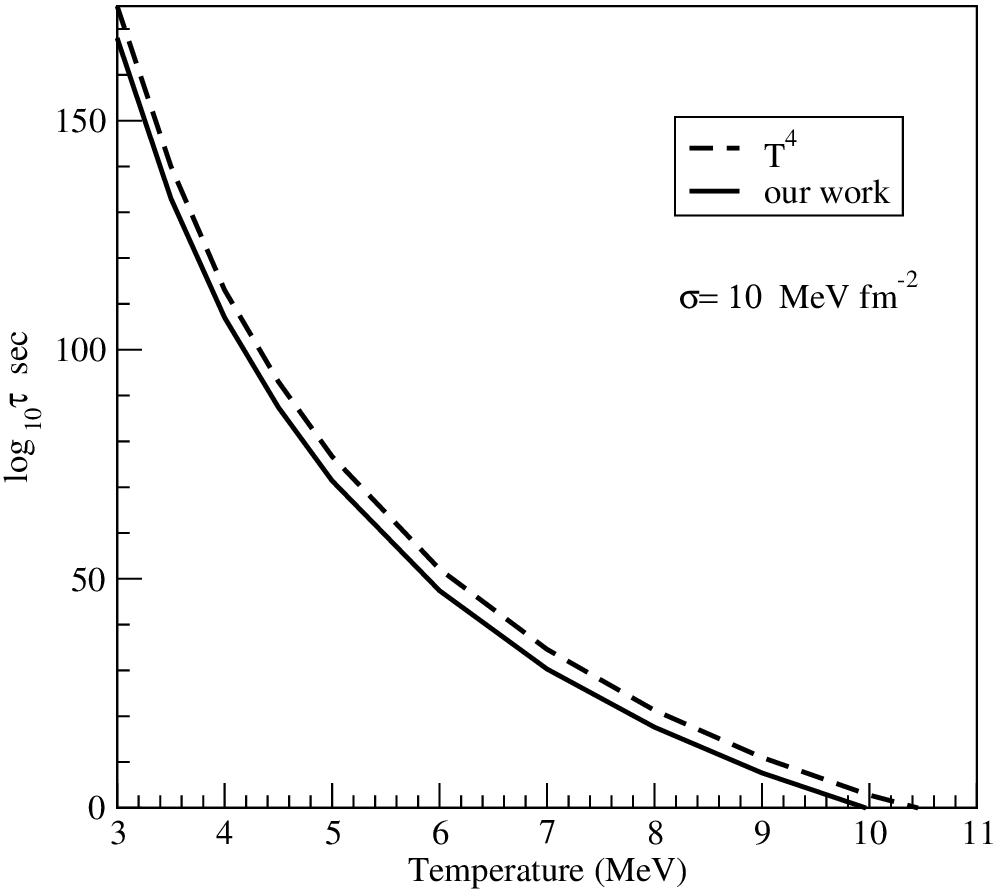}
}}

\vspace{4.0cm}

\noindent{\small{
FIG. 7. Same as Fig.6 but our results are compared with the calculation of 
$T^4$ approximation.}}

\end{document}